\documentclass[twocolumn,amsmath,aps]{revtex4}

\usepackage{epsfig}
\usepackage{subfigure}

\newcommand{\be}{\begin{equation}}
\newcommand{\ee}{\end{equation}}
\newcommand{\bea}{\begin{eqnarray}}
\newcommand{\eea}{\end{eqnarray}}

\begin{document}

\title{An Improved Treatment of Optics in the Lindquist--Wheeler Models}

\author{Timothy Clifton}
\email{tclifton@astro.ox.ac.uk}
\affiliation{Department of Astrophysics, University of Oxford, UK.}

\author{Pedro G. Ferreira}
\email{pgf@astro.ox.ac.uk}
\affiliation{Department of Astrophysics, University of Oxford, UK.}

\author{Kane O'Donnell}
\email{odonnell.kane@gmail.com}
\affiliation{Department of Physics and Astronomy, University of
  Canterbury, New Zealand.}


\begin{abstract}

We consider the optical properties of Lindquist--Wheeler (LW) models
of the Universe.  These models consist of lattices constructed from
regularly arranged discrete masses.  They are akin to the Wigner--Seitz construction of
solid state physics, and result in a dynamical description of the
large--scale Universe in which the global expansion is given by a
Friedmann--like equation.  We show that if these models are constructed
in a particular way then the redshifts of distant objects, as well as
the dynamics of the global space--time, can be made to be in good agreement with the
homogeneous and isotropic Friedmann--Lema\^{i}tre--Robertson--Walker
(FLRW) solutions of Einstein's equations, at the level of $\lesssim 3\%$ out to $z \simeq 2$.
Angular diameter and luminosity distances, on the
other hand, differ from those found in the corresponding FLRW models,
while being consistent with the `empty beam' approximation, together
with the shearing effects due to the nearest masses.  This can be
compared with the large deviations found from the corresponding FLRW values
obtained in a previous study that considered LW models constructed in a
different way.  We therefore advocate the improved LW models we
consider here as useful constructions that appear to faithfully
reproduce both the dynamical and observational properties of
space--times containing discrete masses.

\end{abstract}

\maketitle

\section{Introduction}

In previous papers \cite{CF1,CF2} we have considered luminosity distances and
redshifts in the  (LW) models of the Universe \cite{LW}.
These models treat space--time as a lattice of regular
cells, each with a mass at their center.  The
geometry inside a given cell is then calculated under the
assumption that the influences of all masses external to that cell are
approximately spherically symmetric. This is the direct
gravitational analogue of the Wigner--Seitz construction from 
electromagnetism \cite{WS}.

Unlike the models constructed by Wigner and Seitz, however, the
LW models (without a cosmological constant) are
necessarily dynamical.  This is due to the single charge available in
gravitational theory, that causes a non--zero derivative of the
gravitational field normal to the boundaries of each cell.  The
resulting force then causes the cell boundaries to fall toward, or away
from, the central mass of each cell, and makes the entire lattice
dynamical.  Within this model the phenomenon of an expanding universe
is then an emergent one:  The global cosmological dynamics
result from gluing together the space--times around
individual masses, each of which is well described locally as being static.
These models do not require us to specify a `background geometry' for
the Universe, and therefore provide an opportunity to study the
`back-reaction' of structures on the global evolution of the Universe
\cite{Buchert}.  They also provide a simple model of an inhomogeneous
universe, within which calculations can be readily performed.

Lattices constructed in a positively curved space with no cosmological
constant were considered in the original work on this subject by
Lindquist and Wheeler \cite{LW}. These models were shown to obey
evolution equations identical to those of the spatially homogeneous
and isotropic Friedmann--Lema\^{i}tre--Robertson--Walker (FLRW) solutions
of Einstein's equations with pressureless dust sources.  Only the scale
of the solutions to those equations was found to be different.  This
study was generalized to include lattices constructed in spatially flat
and negatively curved spaces in \cite{CF1,kane}, and to include a
cosmological constant in \cite{CF2}.   Both spatially flat and
negatively curved models were also shown to obey the expected
Friedmann equations.  A further interesting extension
of the original LW models was performed in \cite{uzan},
where lattices with only two cells were considered, and shown to exist 
as solutions of Einstein's equations only when $\Lambda \neq 0$ and
when the masses are out of causal contact with each other.

Unfortunately, the LW models (with the exception of the 2 mass case)
are not exact solutions of
Einstein's equations, and the way that the global lattice is
constructed is not entirely unambiguous.  In their original paper,
Lindquist and Wheeler discussed what they considered to be two
reasonable methods of constructing such a structure; what they
referred to as Condition I and Condition II in \cite{LW} (these
conditions will be discussed in
more detail in Section \ref{model}).  Both of these conditions give
the same Friedmann equations, but lead to different predictions for the
change in scale of the global dynamics.  The optical properties of
spatially flat models that obey Condition II were considered in
\cite{CF1,CF2} \footnote{Note that in the original version of
  \cite{CF1} it was incorrectly stated that the optical properties of
  models obeying Condition I were being studied.}.  In these papers
detailed calculations of the redshifts and luminosity distances that
would be seen by observers in these models were performed, both with
and without a cosmological constant.  Large deviations from the
predictions of the corresponding quantities in FLRW models were found,
and the effects that this could have on cosmological parameter
estimation were considered \cite{CF2}.  This line of study was
generalized to spatially curved models in \cite{kane}.

Here we study the optical properties of spatially flat
Lindquist--Wheeler models using Condition I of \cite{LW}.  It has
recently been shown that the change in scale
predicted by positively curved LW models constructed
using Condition I is in good keeping with the corresponding
exact solutions of Einstein's equations, including back-reaction \cite{C2}.
It has also been shown that spatially flat LW models
constructed using Condition I provide a good description of the global
evolution of space--time when their cells are much smaller than the Hubble 
radius, and much larger than the Schwarzschild radius of the central
masses \cite{C1}.  Condition II does
not provide a good description of the space--time in either of these
cases.  We consider this to provide sufficient motivation to
explore the optical properties of LW models that are
constructed using Condition I, rather than those of the apparently less
favorable Condition II that were studied in \cite{CF1, CF2}.

In Section \ref{model} we briefly describe the way in which the
LW models are constructed, paying particular attention
to the difference between Condition I and Condition II of \cite{LW}.
In Section \ref{obs} we then recap on how observational quantities such as
redshifts and luminosity distances are calculated in these models.  In
Section \ref{results} we show the results of using Condition I when
calculating distance measures and redshifts in the spatially flat models.
It is found that deviations from the corresponding quantities in
spatially flat FLRW models are typically $\lesssim 3\%$ at $z \lesssim
1$, in stark contrast to the results found when using Condition II
\cite{CF1,CF2}.  In Section \ref{disc} we discuss these
results, and their consequences for the validity and applicability of
the LW construction as a model of the Universe.  We conclude that the LW model
provides a surprisingly accurate model of both the cosmological dynamics
(including the back--reaction of structures) and optical properties 
of the space--times associated with regularly spaced discrete masses.
The applicability of these ideas to more irregular configurations of
matter remains to be demonstrated.

\section{The Model}
\label{model}

The LW models are constructed by tiling spaces of
constant curvature with regular polyhedra, and placing a mass $m$ at the
center of each polyhedron.  The geometry of the space--time within each
cell is then calculated under the assumption that the influence of the
external masses is spherically symmetric, and boundary spheres of constant
Schwarzschild radial distance $r$ are allowed to follow geodesics of
the resulting geometry.  This results in the space--time around each
mass being described as Schwarzschild--de Sitter geometry.
By specifying a relation between the boundary spheres and the original
polyhedra one then has a dynamical lattice model of the Universe.

In order to perform calculations one now requires measures of cosmological
size and time.  These are provided in the positively
curved case by the radius of the lattice in an embedding Euclidean
4--space, and by the proper time $\tau$ along the trajectories of the
boundary spheres \cite{LW}.  These concepts were extended to the case
of spatially flat and negatively curved lattices in \cite{CF1}.  The
result is that these lattices obey the evolution equation
\be
\label{Friedmann}
\frac{\dot{a}^2_{LW}}{a^2_{LW}} = \frac{2 m}{a^3_{LW} f^3(\psi)} - \frac{k}{a^2_{LW}},
\ee
where $a_{LW}=r/f(\psi)$ is the size of the lattice (in the original
3--space of constant curvature, to which the trajectories of the
spherical boundaries are made to be orthogonal), $m$ is the `bare mass' at
the center of each cell, $k=1$, $0$ or $-1$ depending on the spatial
curvature of the lattice, and over--dots denote differentiation with
respect to $\tau$.  The function $f(\psi)= \sin (\psi)$, $\psi$ or
$\sinh (\psi)$ for $k=1$, $0$ or $-1$, respectively, where $\psi$ is a
constant (the angle subtended by a single spherical boundary at the origin
of the embedding space when $k=1$).

\begin{figure*}
\vspace{-20pt}
\centering
\subfigure[\; Type I]{
   \includegraphics[height=5.06cm] {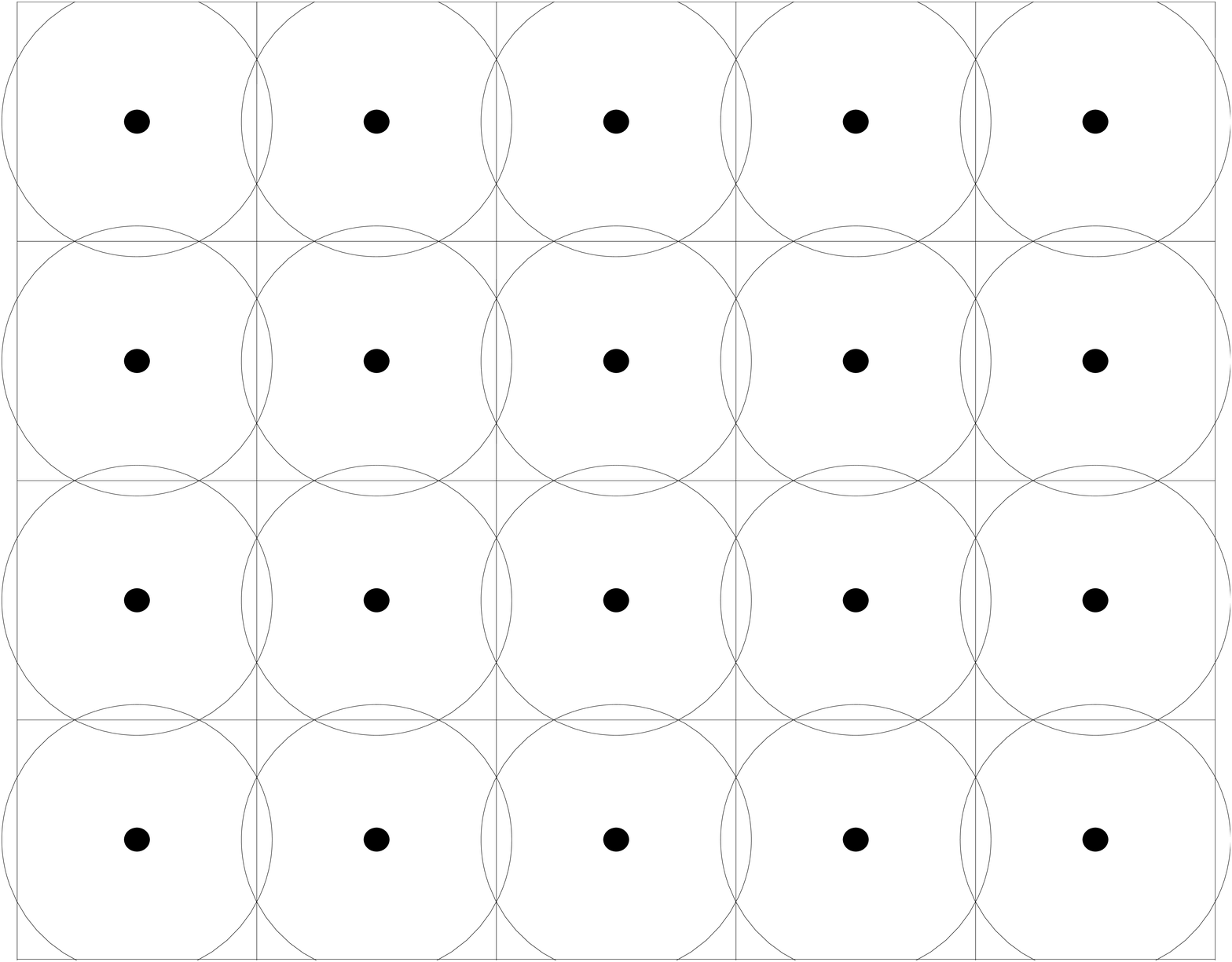}
    }
\hspace{10pt}
 \subfigure[\; Type II]{
   \includegraphics[height=5.08cm] {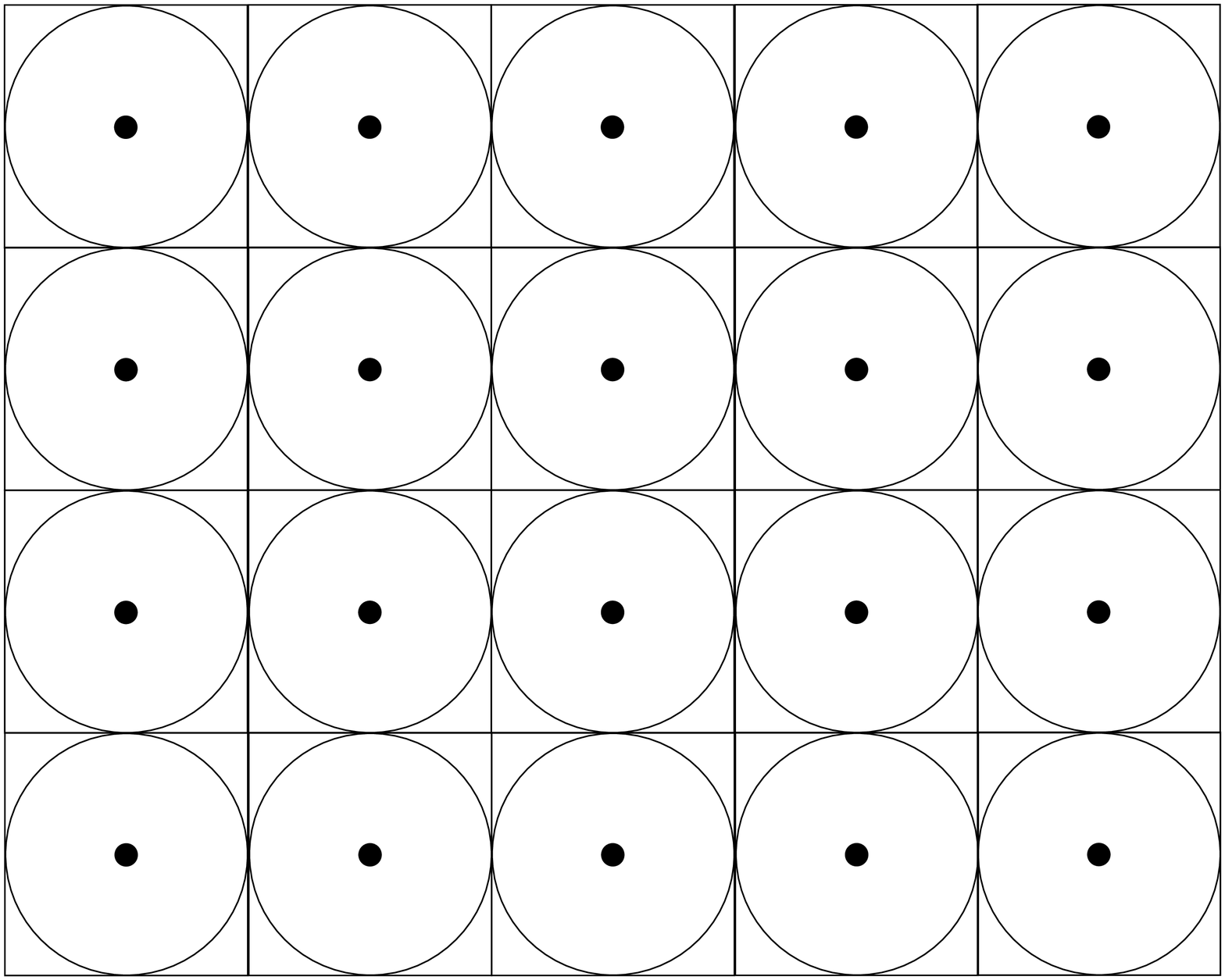}
 }
\caption{Illustration of Type I and Type II lattices, constructed
  using Condition I and Condition II.}
\label{fig1}
\end{figure*}

The precise value of $\psi$ depends on the lattice that is being
constructed (see Appendix A of \cite{CF1} for all the possible tilings
of constant curvature 3--spaces with regular polyhedra), and the way in
which the lattice cells are identified with boundary
spheres. Lindquist and Wheeler suggested two possible conditions for
achieving this:
\newline
\begin{center}
\begin{tabular}{l p{6cm}}
Condition I:  & The boundary sphere shall occupy the same volume of
the lattice as each cell. \\ 
\vspace{-1pt} & \\
Condition II: & The boundary sphere associated with each mass shall
just touch the boundary sphere of all its nearest neighbors.\\
\vspace{-5pt} &  \\
\end{tabular}
\end{center}
These two possibilities are illustrated in Figure \ref{fig1}.  The
value of $\psi$ for each of the six possible lattices that can be
constructed in a positively curved space are then given in Table IV of
\cite{LW}.  In a flat space the only possible lattice has
$\psi=(\frac{6}{\pi})^\frac{1}{3}\simeq 1.24$ under Condition I, and
$\psi=1$ under Condition II.  Throughout this paper we will refer to
lattices constructed using Condition I as Type I, and those
constructed using Condition II as Type II.

It has been shown in \cite{C1} that when the inter--mass separation is
small compared to the Hubble scale, but large compared to the
Schwarzschild radius of the masses, that Type I lattices accurately
reproduce the large--scale dynamics of more accurate solutions to
Einstein's equations.  Using exact solutions, Type I lattices have
also been shown in \cite{C2} to accurately predict the
scale of the maximum of expansion of positively curved universes
(including the predicted deviation from the corresponding FLRW
solutions).  Type II lattices have been shown to produce inaccurate
results for the global dynamics of the space--time in both of
these cases.

\section{Optical Properties}
\label{obs}

The optical properties of spatially flat Type II lattice models were
discussed in \cite{CF1}.  Here we wish to study the optical 
properties of the Type I models, which have been
shown to more faithfully reproduce the correct global dynamics 
\cite{C1,C2}.

The equations of motion for null particles in Schwarzschild--de Sitter
geometry, as appropriate for this study, are \cite{CF1}
\bea
\label{B}
B &=& \left(1- \frac{2 m}{r} -\frac{\Lambda}{3}r^2 \right) \dot{\tau}
+\sqrt{\frac{2 m}{r} +\frac{\Lambda}{3}r^2}\dot{r}\\
\dot{r}^2 &=& B^2 -\frac{J^2}{r^2} \left(1- \frac{2 m}{r}
-\frac{\Lambda}{3}r^2 \right)\\
\dot{\theta}^2 &=& \frac{J^2}{r^4} - \frac{J^2_{\phi}}{r^4 \sin^2
  \theta}\\
\dot{\phi} &=& \frac{J_{\phi}}{r^2 \sin^2\theta},
\label{last}
\eea
where $B$, $J$ and $J_{\phi}$ are constants, over-dots denote 
differentiation with respect to the affine parameter $\lambda$, and
$r$, $\theta$ and $\phi$ are radial and angular Schwarzschild
coordinates.  The remaining quantities $m$ and $\Lambda$ are the bare
mass at the center of each cell and the cosmological constant.

We emphasize that care must be taken in propagating photons
between lattice cells, and that it is not appropriate to simply perform a
translation of the spatial coordinate system.  Rather, one
should ensure that in the rest--space of observers at the cell
boundary the direction of the photon, and the energy of the photon, is unchanged
when leaving the first cell and entering the next.  This can be
achieved by decomposing the tangent vector to the null rays as
\be
k^a = (-u_b k^b) (u^a + n^a),
\ee
where $u^a$ is the tangent vector to the trajectory of an observer at
the boundary, and $n^a$ is a space-like vector that satisfies $n_a
n^a=1$ and $u_an^a=0$.  The energy of a photon that is measured by
this observer is then $E=-u_a k^a$, and the photon direction is given
by the two independent components of $n^a$.  Matching these three
pieces of information at the boundary is sufficient to determine $B$,
$J$ and $J_{\phi}$ in the new cell, which completely specifies the
trajectory of the photon until it hits another cell boundary, at which
point a new matching is performed.

Two methods were proposed in \cite{CF1} for how to choose the $u^a$
that is required for this operation:
\newline
\begin{center}
\begin{tabular}{l p{6.5cm}}
Method I:  & The tangent vector $u^a$ is taken
to be that of the freely falling observers who follow similar curves
to the trajectory of the boundary spheres.\\
\vspace{-1pt} & \\
Method II: & The tangent vector $u^a$ at the cell boundary is taken to be that
of the non--geodesic observers who follow the trajectory of
the cell boundary at the point where the photons cross it. \\
\vspace{-5pt} & \\
\end{tabular}
\end{center}
The first of these two methods is the least computationally demanding,
while the second is constructed to be more accurate.  In what follows
we will therefore present the results of numerical calculations using
Method I, pausing only to make comparisons with Method II in order to
gain an estimate of the systematic errors involved.  A more detailed
description of these two methods is given in \cite{CF1}, to which the
reader is referred for details.

\begin{figure*}[tbh]
\centering
 \subfigure[\; Redshifts along three example trajectories in a Type I
   model, out to $z \sim 0.03$.]{
   \includegraphics[width=7cm] {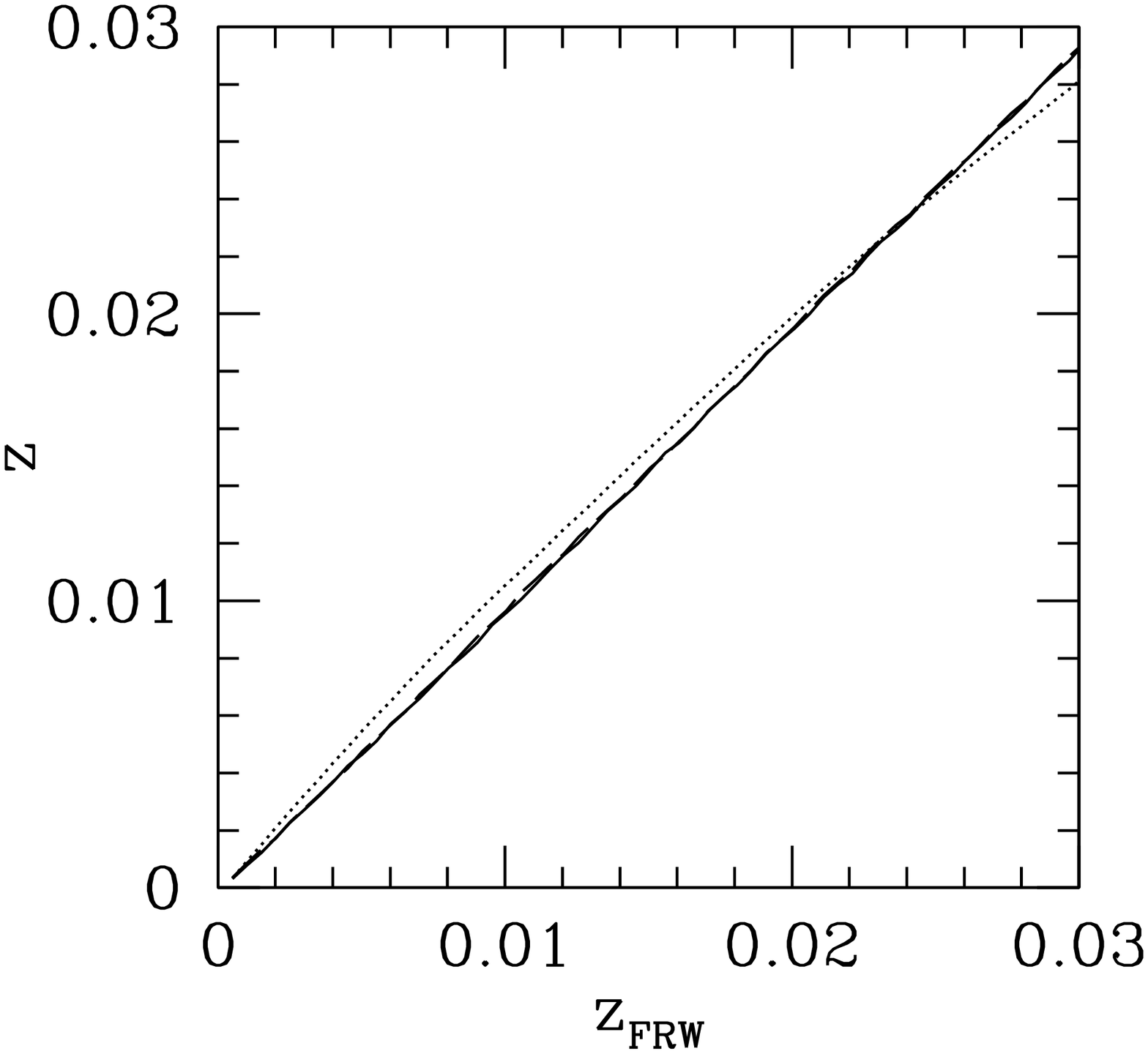}
 }
\subfigure[\; Redshifts in Type I models (dashed line), Type II models
  (dotted line), and FLRW models (solid line).]{
   \includegraphics[width=7cm] {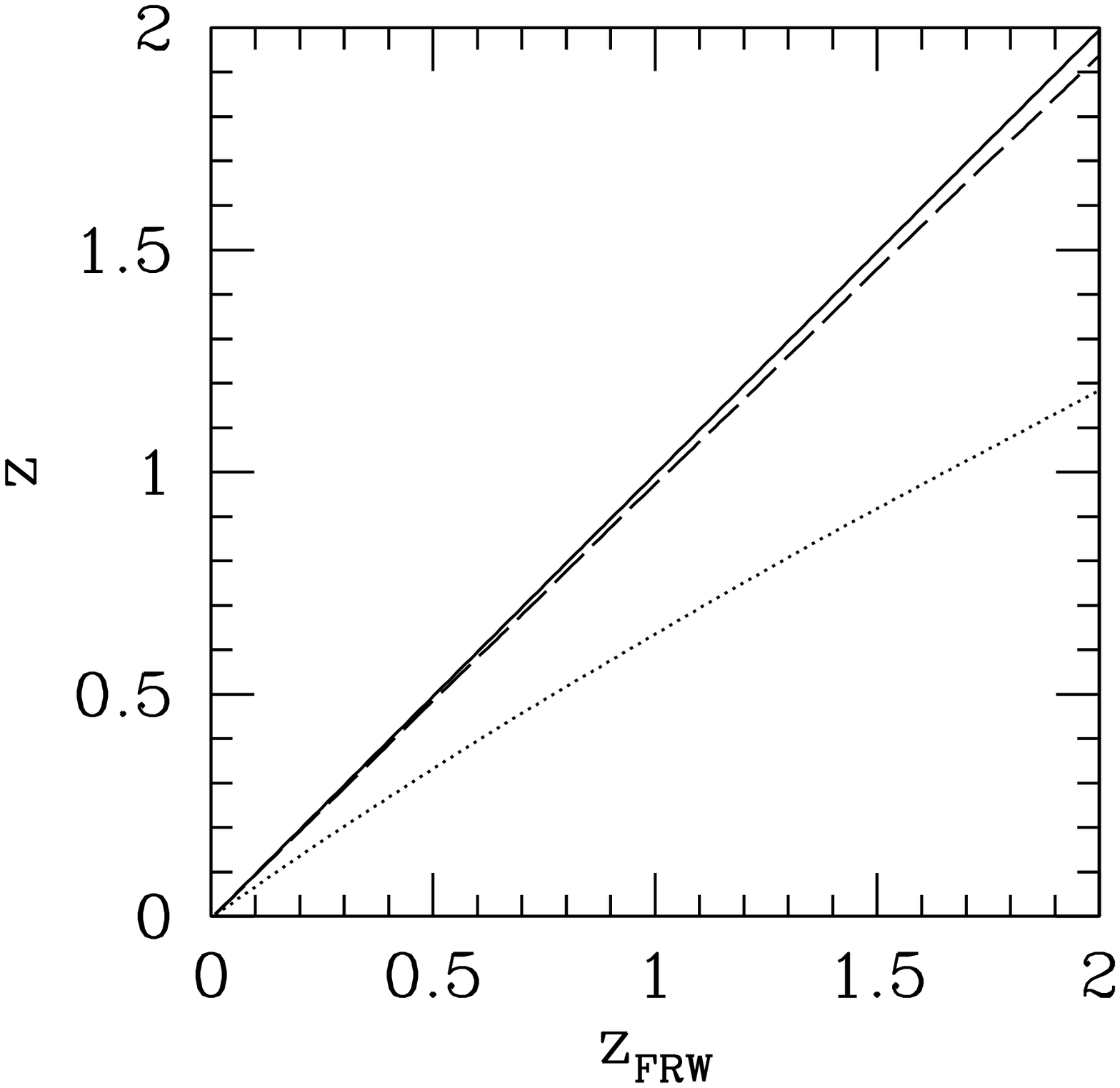}
   }
\caption{Redshifts in spatially flat LW models as a function of the
  corresponding redshift in FLRW cosmology.}
\label{fig2}
\end{figure*}

In order to calculate optical properties along bundles of null
geodesics we use the Sachs equations \cite{sachs}, which in the
present situation can be written as \cite{CF1}
\begin{eqnarray}
\label{S1}
\frac{1}{r_A} \frac{d^2r_A}{d\lambda^2} +\sigma^* \sigma &=& 0\\
\frac{d\sigma}{d\lambda} + \frac{2 \sigma}{r_A} \frac{dr_A}{d\lambda}
&=& \frac{3 m J^2}{r^5} e^{i\Psi},
\label{S2}
\end{eqnarray}
where $r_A$ is the angular diameter distance, $\sigma$ is the complex
shear scalar, and $\Psi$ is the complex phase of the term driving the shear.  As
always, the luminosity distance is then given by $r_L = (1+z)^2 r_A$
\cite{etherington}, and redshifts are given by
\begin{equation}
1+z = \frac{(-u^a k_a)_e}{(-u^a k_a)_o},
\end{equation}
where observers and sources follow curves with tangent vector $u^a$,
photons follow curves with tangent vector $k^a$, and subscripts $e$
and $o$ denote quantities evaluated at the point of emission and
observation of the photons, respectively.  In what follows we take
$u^a$ to be a family of curves that are similar to the trajectory of
the boundary sphere of the lattice cell that they are in.

\section{Results}
\label{results}

Let us now consider the results of performing ray tracing through
models of Type I.  We will first consider cases
with ${\Lambda}=0$, and then cases with ${\Lambda} \neq 0$.

\subsection{Models with $\mathbf{{\Lambda}=0}$}

With $\Omega_{\Lambda}=0$ the space--time inside each cell is described
locally as Schwarzschild geometry.  In this case we can propagate null
trajectories through the lattice and we find that for
Milky Way sized objects separated by $\sim 1$Mpc the resulting
redshifts are as displayed in Figure \ref{fig2}.  The calculated
redshift is shown here as a function of the expected redshift in the
corresponding FLRW model.

The right--hand panel in Figure \ref{fig2} shows redshift, $z$, as a
function of the redshift of objects at the same position in the
corresponding FLRW model, $z_{FLRW}$.  The results of using both Type
I and Type II lattice models are displayed in this plot, as well as
the line $z=z_{FLRW}$, for comparison.  It is clear that the redshifts of
distant objects in LW models is strongly dependent on whether
Condition I or Condition II is applied.  If we parameterize the
redshift in these models as
\begin{equation}
\label{z}
1+z = (1+z_{FLRW})^{\left< \gamma \right>}
\end{equation} 
then we find that the results of using Condition I are well modeled
by  $\left< \gamma \right> \simeq 0.98$, while using Condition II
results in  $\left< \gamma \right> \simeq 0.70$ \cite{CF1}.  This is
a significant difference, with Condition I leading to results
that are in much better keeping with those obtained in standard FLRW.

The left--hand panel in Figure \ref{fig2} shows three individual
realizations of photon trajectories in LW models of Type I.  The direction
of these trajectories is chosen at random, and it can be seen that
while they are not identical, they do oscillate around a mean value.
This mean is rapidly approached as the
distance being observed over is increased, as discussed in detail in
\cite{CF1}.  Over the scales plotted in the right--hand panel of Figure
\ref{fig2} these trajectories cannot be distinguished by eye.
Example trajectories at low $z$ in Type II models are given in \cite{CF1}.

\begin{figure}[t]
\centering
   \includegraphics[width=7cm] {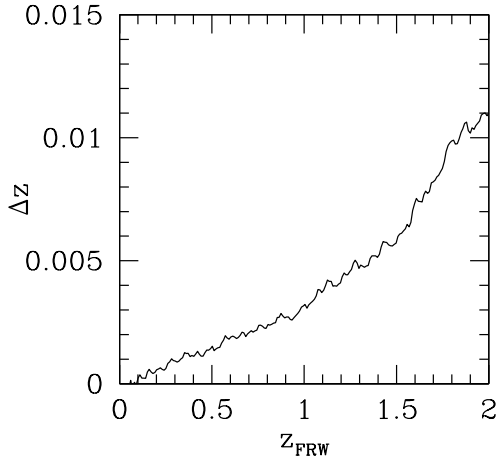}
\caption{The fractional difference between redshifts calculated using
  Method I and Method II in Type I LW models.}
\label{fig3}
\end{figure}

In Figure \ref{fig3} we show the fractional difference in redshifts calculated
using Method I or Method II in models of Type I.  It can
be seen that these two different methods produce 
differences in redshift that are at the level of $\lesssim 1\%$ out to
distances that would correspond to $z\sim 2$ in the corresponding
FLRW models.  This is comparable to the magnitude of the same quantity
calculated in Type II models (shown in Figure 8 of \cite{CF1}), and
provides evidence for the validity of using the computationally less
expensive Method I when calculating observables in these models at low
redshifts.

Integration of the Sachs optical equations in Type I lattice models
proceeds in much the same way as the process discussed in \cite{CF1}
for Type II models.  The shear that
builds in the bundles of null geodesics is still dominated by random
events in which the photon trajectories pass close to the central
masses, and still performs a random walk.  The different condition
used to construct the lattice in the present case does not strongly
affect this conclusion, as the difference between Condition I and
Condition II amounts to a change in the location of the cell
boundary.  The shearing of the bundles of null geodesics, on the other
hand, is dominated by the parts of the photon trajectories that are
closest to the cell centers.  For Milky Way sized masses separated by
$\sim 1$Mpc we find that shear has a negligible affect at
low redshifts in models of Type I, as found for models
of Type II in \cite{CF1}.  The luminosity distance out to $z \sim 1$ in the present
case is then well described by
\be
r_L \sim z + 0.23 z^2 +O(z^3),
\ee
with a corresponding value for the deceleration parameter of 
\be
q_0 \simeq 0.53.
\ee
This is a small deviation from the value of $1/2$ expected in a
spatially--flat, dust--filled FLRW universe with no $\Lambda$, but
considerably less than the value of $q_0 \simeq 1.14$ found in models
of Type II \cite{CF1}.

For further discussion of the effects of shear on distance measures,
and the formation of caustics, the reader is referred to Section 5 of
\cite{CF1}.

\subsection{Models with $\mathbf{{\Lambda} \neq 0}$}

Let us now consider Type I lattice models with $\Omega_{\Lambda} \neq
0$.  Models of Type II with $\Omega_{\Lambda} \neq 0$ were considered
in \cite{CF2}. 

\begin{figure}[b]
\centering
   \includegraphics[width=8.5cm] {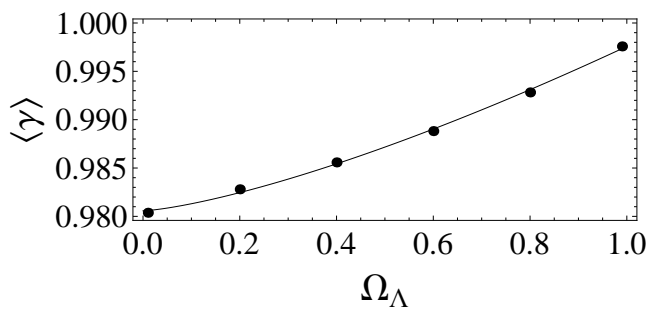}
\caption{$\left< \gamma \right>$ as a function of $\Omega_{\Lambda}$,
  for models of Type I.  These values are much closer to the FLRW
  limit of $\left< \gamma \right>=1$ than the models of Type II.}
\label{fig4}
\end{figure}

In this case the space--time geometry within each cell is approximated
locally as being Schwarzschild--de Sitter.  The first integrals of the
equations of motion for null particles in this geometry are given in
\cite{CF2}, and the propagation of photons between cells proceeds in
a similar fashion to the $\Omega_{\Lambda}=0$ case.  The redshift in
these situations can again be well modeled by Equation (\ref{z}), but
where the new parameter is now given by $\left<\gamma\right> =
0.98+0.017 \Omega_{\Lambda}^{1.4}$.  This function is displayed
graphically in Figure \ref{fig4}.  The corresponding quantity for Type
II models can be found in \cite{CF2}.

Using the redshift relation specified by Equation \ref{z} and Figure
\ref{fig4} we can now calculate distance measures as functions of
redshift, as outlined in Section \ref{obs}.  We will be particularly
interested in the distance modulus, which is a useful quantity to
compare to supernova observations.  This is defined as
\be
\Delta {\rm dm} = 5 \log_{10} \left( \frac{r_A}{r^{\rm Milne}_A} \right),
\ee
where $r^{\rm Milne}_A$ is the angular diameter distance in an open, empty Milne
universe with the same local Hubble rate.  This definition is equivalent to the magnitude of a source,
minus the magnitude it would have at the same redshift in a Milne universe.

\begin{figure*}[t]
\centering
   \includegraphics[width=16.8cm] {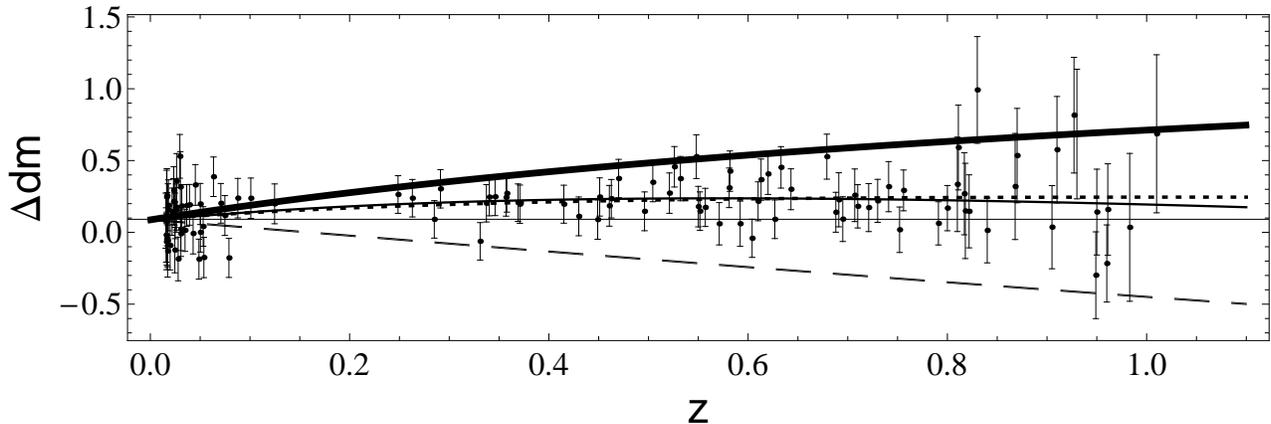}
\caption{The best fitting spatially flat LW model of Type I
  (dotted line), and the best fitting spatially flat FLRW model (thin
  line).  The data represents 115 high and low redshift supernovae
  from SNLS \cite{SNLS}, fitted to the LW model
using the SALT light curve fitter \cite{SALT}. Einstein-de Sitter
(dashed line) and de Sitter (thick line) models are also shown, for
reference.}
\label{fig5}
\end{figure*}

Fitting the LW models of Type I to the SNLS supernova data \cite{SNLS}
we find that the best fitting spatially flat model has
$\Omega_{\Lambda}=0.66 \pm 0.05$, while the best fitting spatially
flat FLRW model to the same data set has $\Omega_{\Lambda}=0.74 \pm
0.04$.  The distance modulus curves for these two models are displayed in Figure
\ref{fig5}.  The corresponding best fit value using Condition II is
$\Omega_{\Lambda} = 0.66 \pm 0.04$ \cite{CF2}.  Unlike the redshift
relations, we therefore find that the
difference in the best fitting value of $\Omega_{\Lambda}$ between LW
models constructed using Condition I and Condition II is not large.
The fit to the data using Condition I is marginally worse
than that found using Condition II, however, with a difference in log
likelihood of $\vert \Delta \ln \mathcal{L} \vert \simeq 0.1$ between them.

\section{Discussion}
\label{disc}

In this paper we have considered the optical properties of
the Lindquist--Wheeler cosmological models constructed from a lattice
of discrete masses.
This study builds on previous work that found large deviations from
the expected values derived from models that contain a continuous
distribution of matter \cite{CF1,CF2}.

We find that the way in which the lattice is constructed has
considerable consequences for its optical properties.  In particular,
using Condition I of \cite{LW} to construct
lattice models results in redshifts that are within a few percent of
the expected values from the corresponding FLRW models.  This should
be compared with differences of $\sim 1/3$ at redshifts of $z\sim 1$
that were found using Condition II of \cite{LW}.  Other aspects of the
optical properties of these models, however, are less sensitive to the
way in which the lattice is constructed.  The accumulation of shear
along bundles of null geodesics, for example, is largely unaffected.
Interestingly, we also find that when $\Lambda$ is included in these
models the best fitting value of $\Omega_{\Lambda}$ to supernova data is largely
insensitive to the whether we use Condition I or II.  In
either case $\Omega_{\Lambda}$ is $\sim 10\%$ less than is required in
FLRW cosmology, in order to fit the same data.

Recent studies of cosmological solutions to Einstein's equations that
contain discrete masses, rather than a continuous fluid, suggest that
the Lindquist--Wheeler models constructed using Condition I provide a
good description of the large--scale evolution of a universe filled
with regularly arranged discrete masses \cite{C1, C2}.  It even
appears that they provide a reasonably good
description of the back--reaction that such structures should have on
the large--scale evolution of an initially homogeneous and isotropic
space (this effect being large for a small number of very large
masses, and small when very many small masses are considered).
The present study supports these results by showing
that LW lattice models can also accurately reproduce the expected
optical properties of a universe filled with many low-mass, discrete,
regularly arranged masses \cite{C1}.  

Collectively, this evidence suggests to us that the LW models constructed using
Condition I provide a simple and reasonably accurate way to model very
highly inhomogeneous distributions of matter \footnote{This can be
  contrasted with the LW models constructed using Condition II, which
  have now been shown to inaccurately reproduce both the required dynamics
  \cite{C1,C2} and the required optics \cite{C1}.}.  They do this without
{\it a priori} assuming any background geometry for the Universe, and
have been shown to give information that cannot be obtained from
exact FLRW solutions alone \cite{C2}.  The LW models therefore appear
to us to be a promising way to investigate the possible effects of
non--linear density fluctuations in relativistic cosmology.

\section*{Acknowledgments}

TC and PGF acknowledge the support of the STFC, the BIPAC, and the
Oxford Martin School.  We are grateful to George Ellis, Kjell Rosquist
and Reza Tavakol for helpful discussions and suggestions.

\end{document}